\documentstyle[preprint,eqsecnum,aps]{revtex}

\begin{document}
\draft
\title{A Relativistic Calculation of Super-Hubble Suppression of
  Inflation with Thermal Dissipation}
\author{Wolung Lee$^{1,2}$ and Li-Zhi Fang$^1$}

\address{$^1$ Department of Physics, University of Arizona, Tucson, AZ 85721}
\address{$^2$ Institute of Physics, Academia Sinica, Taipei, Taiwan 11529,
R.O.C.}
\date{\today}
\maketitle
\begin{abstract}

We investigated the evolution of the primordial density perturbations
produced by inflation with thermal dissipation. A full relativistic analysis
on the evolution of initial perturbations from the warm inflation era to a
radiation-dominated universe has been developed. The emphasis is on tracking
the ratio between the adiabatic and the isocurvature mode of the initial
perturbations. This result is employed to calculate a testable factor: the
super-Hubble suppression of the power spectrum of the primordial
perturbations. We show that based on the warm inflation scenario, the
super-Hubble suppression factor, $s$, for an inflation with thermal
dissipation is at least $0.5$. This prediction does not depend on the details
of the model parameters. If $s$ is larger than $0.5$, it implies that the
friction parameter $\Gamma$ is larger than the Hubble expansion parameter $H$
during the inflation era.

\end{abstract}


\narrowtext

\section{Introduction}
\label{sec:level1}

According to the standard inflationary cosmology \cite{KT}, the formation of
structure in the universe starts from small amplitude density perturbations
generated by the fluctuations of scalar fields during the inflation epoch.
The initial perturbations then subsequently develop through gravitational
instability. Such a mechanism generally produces adiabatic
perturbations with a roughly scale-invariant fluctuation spectrum possessing
an index $\sim 1\ $\cite{KT}. That is, the power spectrum of the primordial 
density perturbations can be written as
\begin{equation}
P(k) \propto k^n,
\end{equation}
where $k$ is the comoving wavenumber, and $n\simeq 1$. The tiny temperature
anisotropies observed in the cosmic microwave background (CMB) are basically
consistent with this scenario\cite{dmr4}.

However, the first two years of the Cosmic Microwave Explorer (COBE) 
Differential Microwave Radiometer (DMR) observations on the CMB temperature 
fluctuation have found the lack of perturbations on scales larger than 
$\simeq 1/H_0$, where $H_0$ is the Hubble constant \cite{jf}. Namely, 
there is a lack of power in the spectrum of the initial density 
perturbations on length scales equal to or larger than the Hubble radius 
$1/H_0$.

More recently, both $\chi^2$ and likelihood analyses of the COBE-DMR
2-year and 4-year sky maps found that the CMB temperature fluctuations
can be fitted by models with power spectrum as
\begin{equation}
P(k) \propto \frac{k^n}{1+(k_{min}/k)^m},
\end{equation}
where $m\geq 4$, and $k_{min} \simeq H_0$\cite{fj}. Obviously, the factor 
$1/[1+(k_{min}/k)^4]$ implies the suppression of density perturbations
 on scales larger than Hubble radius, $1/k_{min}$.

The super-Hubble suppression is difficult to conform with the purely
adiabatic (isentropic) initial perturbations of scalar field $\phi$ of the
standard inflation. In the framework of inflation, a possible mechanism of
the super-Hubble suppression is the multi-component inflation scenarios.
The initial perturbations produced by such models are hybrid, i.e. they
contain both adiabatic and isocurvature fluctuations \cite{kps}. The
behaviors of adiabatic and isocurvature perturbations are qualitatively
different on super-horizon scales. Principally, isocurvature fluctuations
in the super-horizon region will give rise to isothermal perturbations,
which do not contribute to density perturbations on super-Hubble scales.
Yet, when the isothermal perturbations enter the Hubble horizon, they will
become density perturbations. Thus, if the initial perturbations are hybrid,
the perturbations shall show an amplitude enhancement at scale $\sim 1/H_0$.
In other words, the hybrid perturbations at super-Hubble scales are
suppressed. 

In the current pool of inflation models, many of them are of multi-component.
It includes double and variant multiple inflation\cite{ps}, bulk viscous
cosmology\cite{zj}, hidden sector of supersymmetry inflation\cite{rs}, etc.
However, the initial entropy fluctuations may not survive or significantly 
be reduced due to the post-inflationary reheating which is essentially a 
process of entropy production and thermalization. Therefore, multi-component 
inflations may not be sufficient to explain, or to quantify the super-Hubble 
suppression if a reheating is necessary. Accordingly, to settle the issue 
of the super-Hubble suppression, a multi-component inflation without 
reheating is preferred.

 From the studies of the system consisting of a scalar field and thermal
bath, it has been realized that two components, the scalar field and the
thermal bath, can coexist during inflation\cite{bf}. Therefore, the mass
density perturbations from the inflation with thermal dissipation should be
hybrid perturbations. Moreover, analytic and numerical solutions of this
model illustrated that the radiation component increases continuously and
smoothly during the inflation epoch, and there is no post-inflationary
reheating\cite{lf1,lf2}. Therefore, the hybrid perturbations generated during
inflation can survive the cosmic inflation and imprint on the subsequent
evolution.

Recently it was found that this model can fit in with the amplitude and index
of the power spectrum of the observed CMB temperature fluctuations if the
mass of inflaton, $M$, is taken to be $\sim 10^{15}-10^{16}$ GeV\cite{lf2}.
Thus, it is worth examining whether the observed supper-Hubble suppression
also corresponds to this model with the same parameters.

The goal of this paper is to calculate the super-Hubble evolution of
hybrid perturbations, and then the super-Hubble suppression for inflation 
model with significant thermal dissipation.
Because the super-Hubble or super-horizon evolution is relativistic in
nature, the super-Hubble suppression must be obtained using a full
relativistic analysis.

This paper is organized as follows. In Sec. II we briefly introduce the model
of inflation with thermal dissipation. The relativistic perturbation
calculations are given in Sec. III. Section IV presents the evolution of
the adiabatic and isocurvature modes of the density perturbations and obtains
the super-Hubble suppression according to the hybrid perturbations. We then
summarize our findings in Sec. V. 

\section{Inflation with Thermal Dissipation}

Let us briefly introduce the model of inflation with thermal dissipation.
The details have been described in \cite{bf,lf1,lf2}. Consider a flat
universe consisting of a scalar field $\phi$, the inflaton, and a thermal
bath. The equations of the expanding universe are given by
\begin{equation}
\dot{H} = -4\pi G\left(\dot{\phi}^2+\gamma\rho_r\right),
\end{equation}
and
\begin{equation}
H^{2} = \frac{8\pi G}{3}\left[\rho_{r} +
\frac{1}{2}\dot\phi^{2} + V(\phi)\right ],
\end{equation}
where $\gamma=4/3$ represents the adiabatic index of thermal radiation, 
$H=\dot{a}/a$ is the Hubble parameter, and $a$ is the cosmic scale factor. 
The effective potential for the $\phi$-field is denoted by $V(\phi)$, and
$\rho_{r}$ is the energy density of the thermal bath. 

The equation of motion for the scalar field $\phi$ is \cite{rey}
\begin{equation}
\ddot{\phi} + 3H \dot{\phi} + \Gamma\dot{\phi} 
- e^{-2Ht}\nabla^2\phi+ V_{,\phi}(\phi)=0.
\end{equation}
The induction of the friction term $\Gamma \dot{\phi}$ represents a
possible approximation for the dissipation of $\phi$ field in a heat bath
environment in near-equilibrium circumstances.  Accordingly it describes
the interaction between the $\phi$ field and the thermal component.  In
principle, $\Gamma$ can be a function of $\phi$. In the cases of
polynomial interactions between $\phi$ field and bath environment,
one may take the polynomial of $\phi$ for $\Gamma$, i.e.
$\Gamma= \Gamma_m \phi^m$ \cite{OR}. The friction coefficient must be
positive definite, hence $\Gamma_m > 0$, and the dissipative index of
friction $m$ should be zero or even integer if $V(\phi)$ is invariant
under the transformation $\phi \rightarrow -\phi$.

For a uniform field, or the background $\phi$, the term $\nabla^2\phi$ in
Eq. (2.3) can be ignored, and we have
\begin{equation}
\ddot{\phi} + (3H + \Gamma) \dot{\phi} + V_{,\phi}(\phi)=0.
\end{equation}
In the case of slow-rolling, Eq.(2.4) yields
\begin{equation}
\dot{\phi} \simeq - \frac{V_{,\phi}(\phi)}{3H + \Gamma}.
\end{equation}

The equation of the radiation component (thermal bath) is given by the first
law of thermodynamics as
\begin{equation}
\dot \rho_{r} + 3 H \gamma \rho_{r} = \Gamma\dot{\phi}^{2}.
\end{equation}
The temperature of the thermal bath is related to the radiation energy 
density by $\rho_{r}=(\pi^{2}/30) g_{\rm eff}T^{4}$, where $g_{\rm eff}$ is 
the effective number of degrees of freedom at temperature $T$. 

The warm inflation epoch is defined when the temperature
of the thermal bath is larger than the Hawking temperature, i.e.
\begin{equation}
T > H.
\end{equation}
Since the thermal and the quantum fluctuations of the scalar field $\phi$ are
proportional to $T$ and $H$ respectively, Eq. (2.7) indicates that during the
warm inflation epoch, the thermal fluctuations dominate over the quantum
fluctuations.

Equation (2.7) is also necessary for maintaining the thermal equilibrium of
the radiation component. In general, the time scale for the relaxation of a
radiation bath is shorter for higher temperature. Accordingly, to have the
relaxation time of the bath be shorter than the expansion time scale of the
universe, a temperature higher than $H$ is needed.

For a $\phi^4$ potential, e.g.  
$V(\phi)=\lambda(\phi^{2}-\sigma_M^{2})^{2}$ with $M^4\equiv V(0)=\lambda
\sigma_M^4$ , the background solution of the $\phi$ field is $\phi =\phi_i
e^{\alpha_M Ht}$, where $\alpha_M\simeq\lambda^{1/2}(m_{\rm
Pl}/M)^2/2\pi$, the Planck mass $m_{\rm Pl}=1/\sqrt{G}$ and $\phi_i$ is
the initial value of the scalar field. The temperature evolution is given
by $T(t) \simeq He^{\alpha_M H (t-t_e)/2}$, which is larger than $H$ when
$t > t_e$. Thus, the kinetic energy density of $\phi$ field and radiation
energy density are, respectively
\begin{equation}
\frac{1}{2}\dot{\phi}(t)^2 \simeq \lambda^{1/2}
\left(\frac{m_{\rm Pl}}{M} \right )^2
\left (\frac{\phi(t)}{\sigma}\right)^2 V(0),
\end{equation}
\begin{equation}
\rho_r(t) \simeq
\lambda^{1/2} \left(\frac{m_{\rm Pl}}{M} \right )^2 \frac{\Gamma}{H}
\left (\frac{\phi(t)}{\sigma}\right)^2 V(0).
\end{equation}
It has been shown that the background solutions are allowed for a very
wide parameter range, as $10^{-7}H < \Gamma < 3H$\cite{lf2,OR}. 

A typical background solution of $H$ and $T$ is shown in Fig. 1. 
The Hubble parameter, $H(t)$ [Eq. (2.2)], remains  constant during the 
inflation. This is the same as in the standard slow-roll inflation model. 
At the end of inflation $t_f$, the Hubble parameter $H(t)$ evolves from 
$H(t) \sim $ constant to a radiation regime $H(t) \propto t^{-1}$. 
In the range of $T < H$, i.e. $t<t_e$, the solution is  unphysical, 
because  it is impossible to maintain the thermalization of
the heat bath with the temperature less than the Hawking temperature $H$.
Thus, the warm inflation epoch is specified by the period of $t_e < t <
t_f$.

Figure 1 shows that the inflation smoothly exits to a radiation era at $t_f$,
i.e. $\rho_{r} \sim V(0)$ for $t>t_f$. There is no reheating epoch in this
model. One can calculate the evolution of density perturbations from the
slow-rolling era to radiation era without the interruption of reheating. This
is especially important for the primordial entropic or isocurvature
fluctuations because the post-inflationary reheating is generally a
thermalization process with a time scale less than $1/H$, thus it may 
smooth away initial entropic fluctuations, and even the initial 
isocurvature perturbation cannot survive the reheating.

\section{Perturbations during warm inflation}

\subsection{Equations of a perturbed universe}

To describe a perturbed universe, we employ the so-called gauge-specific
formalism \cite{b80,b88,h123}. The matter distribution is described by a
normalized 4-velocity vector $u^a$ at each point of space-time, i.e.
$u^au_a=-1$ (the Latin indices run from 1 to 4, and the Greek indices from
1 to 3). At each point, a projection into the rest space of an observer
moving with $u^a$ can be made by a projection tensor defined as
$h_{ab}\equiv g_{ab}+u_au_b$. As such, the line element of the spacetime
can be written as\cite{el}
\begin{equation}
ds^2\equiv g_{ab}dx^adx^b=-\left(u_adx^a\right)^2+h_{ab}dx^adx^b.
\end{equation}
An observer at the space-time point $x^a$ moving with 4-velocity $u^a$
assigns to the event $x^a+dx^a$ a spatial separation $(h_{ab}dx^adx^b)$,
and a time separation $|u_adx^a|$ from him. 

The covariant derivative of the 4-velocity $u^a$ can be decomposed into 
the vorticity tensor $\omega_{ab}$, the shear tensor $\sigma_{ab}$, the 
expansion scalar $\theta$, and the acceleration $\dot{u}_a$ as
\begin{equation}
u_{a;b}=\omega_{ab}+\sigma_{ab}+\frac{1}{3}\theta h_{ab}-\dot{u}_au_b,
\end{equation}
where $\sigma^a_{~a}=\sigma_{ab}u^b=\omega_{ab}u^b=\dot{u}_au^a=0$, with
$\sigma^2\equiv\frac{1}{2}\sigma_{ab}\sigma^{ab},\ 
\omega^2\equiv\frac{1}{2}\omega_{ab}\omega^{ab}$, and 
$\theta\equiv u^a_{~;a}$. The overdot here denotes a covariant derivative 
along the $u^a$ direction.

The energy-momentum tensor of matter can generally be written as
\begin{equation}
T_{ab}=\rho u_au_b+ph_{ab}+q_au_b+q_bu_a+\pi_{ab},
\end{equation}
where $q_au^a=\pi_{ab}u^b=0$ and $\pi_{ab}=\pi_{ba}$.  The total energy
density of matter measured by observer $u^a$ is represented by $\rho$, while
$p$ and $\pi_{ab}$ denote the isotropic and anisotropic pressures
respectively. The quantity $q_a$ prescribes the energy flux relative to
$u^a$.

There are two components of matter in the era of warm inflation: a scalar
field and a thermal bath. Thus, the total energy-momentum tensor is
\begin{equation}
T_{ab}=T_{(\phi)ab} + T_{(r)ab},
\end{equation} 
where subscripts $(\phi)$ and $(r)$ are for the scalar field and the thermal
bath, respectively. The energy-momentum conservation is given by 
\begin{equation}
T^{ab}_{~~;b}=0,
\end{equation}
and 
\begin{equation}
T_{(\phi)a;b}^{~~b}\equiv Q_{(\phi)a}, \hspace{6mm}
T_{(r)a;b}^{~~b}\equiv Q_{(r)a}, 
\end{equation}
where $Q_{(\phi)a}$ and $Q_{(r)a}$ describe the interaction between 
$\phi$ field and the thermal bath. From Eqs. (3.4), (3.5) and (3.6), we 
have
\begin{equation}
Q_{(\phi)a} + Q_{(r)a} =0.
\end{equation}

The interaction term $Q_{(i)a}$ ($i$ is for $\phi$ or $r$) can be further 
decomposed into \cite{ks}
\begin{equation}
Q_{(i)a}\equiv Q_{i}u_{(i)a}+J_{(i)a} \hspace{6mm} {\rm with} \hspace{6mm}
u_{(i)}^{~a}J_{(i)a}=0~.
\end{equation}
The quantities $Q_{i}$ and $J_{(i)a}$ are, respectively, the temporal and
the spatial components of $Q_{(i)a}$. The spatial components of $J_{(i)a}$
can be defined as $J_{(i)\alpha}\equiv J_{i,\alpha}$ in general\cite{h123}.
According to Eq. (3.7), one has
\begin{equation}
Q_{\phi} +Q_{r} =0 \hspace{6mm} {\rm and} \hspace{6mm}
J_{\phi} +J_{r} =0.
\end{equation}
 From Eq. (3.6), we have $-u_{(i)}^aT_{(i)a;b}^b=Q_{i}$ and
$h_{(i)c}^aT_{(i)a;b}^b= J_{(i)c}$. 

For dissipation models, we need to express the interaction 
$Q_{(i)a}$ in terms of the frictional coefficient $\Gamma$. From Eqs.
(2.4) and (2.6), $Q_{(i)a}$ in the background are described by solutions 
\begin{equation}
Q_{\phi} = -Q_r = -\Gamma\dot{\phi}^2,
\end{equation}
and
\begin{equation}
J_{\phi} = -J_r = 0.
\end{equation}
Therefore, the general expression of $Q_{(i)a}$ can be found by 
making (3.10) and (3.11) explicitly covariant. We have then
\begin{eqnarray}
Q_{\phi} & = &  -\Gamma (\phi_{,a}u^a)^2 = -u_{(\phi)}^aT_{(\phi)a;b}^b
   \\ \nonumber
Q_r & = &  \Gamma (\phi_{,a}u^a)^2 = -u_{(r)}^aT_{(r)a;b}^b 
\end{eqnarray}
and 
\begin{equation}
h_c^aT_{(\phi)a;b}^b = - h_c^aT_{(r)a;b}^b =
  - \Gamma (\phi_{,b}u^b)\phi_{,a}h_{c}^a.
\end{equation}
 
\subsection{The variables of perturbations}

\subsubsection{Perturbation variables of space-time metric}

For a uniform and isotropic background, the space-time of the universe is
described by a flat Friedman-Robertson-Walker metric as
\begin{equation}
ds^2=\bar{g}_{ab}dx^adx^b= -(u_0)^2dt^2+a(t)^2\left[dr^2+
r^2\left(d\theta^2+\sin^2\theta d\phi^2\right)\right],
\end{equation}
where $u^0$ is the temporal 4-vector of $u^a$. During inflation,
the cosmic scale factor $a(t)=e^{Ht}$.
The perturbed metric is prescribed by $g_{ab}=\bar{g}_{ab}+\delta g_{ab}$.
In this paper, we will be
interested in the scalar-type perturbations only. In this case, the
perturbations of the background metric can be described by four metric
variables\cite{b88}
\begin{eqnarray}
{\rm lapse \ function \ \alpha:} & g_{00}=-a^2(1+2\alpha) \\
  \nonumber
{\rm  3-space \  curvature \ \varphi:} &
{\cal R}=\frac{4k^2}{a^2}\varphi \\ \nonumber
{\rm expansion \ scalar \  \kappa:} & \theta=3H-\kappa \\
   \nonumber
{\rm  shear \ \chi:} & \ \ 
\sigma_{\alpha\beta}=\chi_{\left.\right|\alpha\beta}-\frac{1}{3}
g^{(3)}_{\alpha\beta}\chi^{\left.\right|\gamma}_{~\left.\right|\gamma} \\
  \nonumber
\end{eqnarray}
where the comoving spatial metric tensor $g^{(3)}_{\alpha\beta}$ satisfies
$\bar{g}_{\alpha\beta}=a^2g^{(3)}_{\alpha\beta}$. The vertical bar
indicates a covariant derivative based on $g^{(3)}_{ab}$.

\subsubsection{Perturbation variables of radiation}

The thermal bath or the radiation component can be treated as a
relativistic ideal fluid. For the background solution, the energy flux
$q_{\alpha}=0$, and the anisotropic pressure $\pi_{ab}=0$. The perturbations
of the energy-momentum tensor of the radiation
$T_{(r)ab}=\overline{T}_{(r)ab}+\delta T_{(r)ab}$
can be described by four matter variables defined as: 
\begin{eqnarray}
{\rm perturbation \ of \  the \ energy \ density:} &  
   \varepsilon_{r} \equiv \delta\rho_{r}\  \\ \nonumber
{\rm perturbation \ of \ isotropic \ pressure:} &
    \pi_{r}\equiv\delta p_r = 
(\gamma-1)\varepsilon_r \  \\ \nonumber
{\rm energy \ flux:}   & q_{\alpha}\equiv\psi_{{r},\alpha}\ \\ \nonumber
{\rm anisotropic \ pressure:} &  \ \ 
\pi_{\alpha\beta}\equiv \sigma_{\left.\right|\alpha\beta}-\frac{1}{3}
g^{(3)}_{\alpha\beta}\sigma^{\left.\right|\gamma}_{~\left.\right|\gamma} \\
  \nonumber
\end{eqnarray}
where $\psi_{r}$ is the comoving velocity of the radiation component. It is
reasonable to take $\sigma = 0$ for radiation.

\subsubsection{Perturbation variables of the scalar field}

For a minimal coupled scalar field $\phi$, the fluid quantities in the
energy-momentum tensor (3.3) are given by
\begin{equation}
\rho_{\phi} = \frac{1}{2}(\phi,_a u^a)^2+V(\phi), \hspace{3mm}
p_{\phi} = \frac{1}{2}(\phi,_a u^a)^2-V(\phi), \hspace{3mm}
q_a=-h^b_a\phi_{,b}\phi_{,c}u^c, \hspace{3mm} \pi_{ab}=0.
\end{equation}
Since the background solution of the $\phi$ field is spatially uniform,
one has $\phi_{,\alpha}=0$, and therefore,  $\rho_{\phi} =
\dot{\phi}^2/2+V(\phi)$, $p_{\phi}=\dot{\phi}^2/2-V(\phi)$,
$q_a=0$ and $ \pi_{ab}=0$.
                                                
Perturbing the term $\dot{\phi}=\phi,_a u^a$, we obtain
\begin{equation} 
\delta \dot{\phi}=\delta\phi,_a u^a+ \phi,_a\delta u^a =
\delta\phi,_0u^0+\phi,_0 u^0\left(\frac{\delta u^0}{u^0} \right ).
\end{equation}
By means of (3.15), we have $\delta u^0/u^0 \simeq 
-\delta g_{00}/ 2g_{00} = -\alpha$. Therefore, to linear order, Eq. (3.18) 
gives
\begin{equation}
\delta (\phi,_a u^a) = \delta\dot{\phi}-\dot{\phi}\alpha,
\end{equation}
and 
\begin{equation}
\delta (\phi,_a u^a)^2 = 2\dot{\phi}\delta\dot{\phi}-2\dot{\phi}^2\alpha.
\end{equation}
Thus, the perturbed energy density of the scalar field $\phi$ is 
given by
\begin{equation}
\varepsilon_{\phi} \equiv \delta\rho_{\phi} =
 \dot{\phi}\delta\dot{\phi}-\dot{\phi}^2\alpha+V,_{\phi}\delta\phi.
\end{equation}
Similarly, the perturbation in the pressure of the $\phi$ field is 
\begin{equation}
\pi_{\phi} \equiv \delta p_{\phi} = 
\dot{\phi}\delta\dot{\phi}-\dot{\phi}^2\alpha-V,_{\phi}\delta\phi.
\end{equation}

The internal energy flux of the perturbed $\phi$ field,
$q_{\alpha}=\psi_{\phi,\alpha}$ is given by
\begin{equation}
\psi_{\phi}=-\dot{\phi}\delta\phi.
\end{equation}
Since $ \pi_{ab}=0 $, the perturbed anisotropic pressure
$\sigma_{\phi}=0$.

\subsubsection{Perturbations of interaction terms}

If we perturb the temporal component of the interaction term upon
the background to linear order, Eq. (3.12) gives rise to
\begin{equation}
\delta Q_r=-\delta Q_{\phi}= \delta [ \Gamma (\phi_{,a}u^a)^2]  =
 2\Gamma(\dot{\phi}\delta\dot{\phi}-\dot{\phi}^2\alpha),
\end{equation}
where Eq. (3.20) has been used. We have taken 
$\delta \Gamma =0$, i.e. assuming $m=0$ in the model 
$\Gamma = \Gamma_m\phi^m$, because the mass power spectrum of the
$\phi$ field is found to be rather insensitive to the dissipative index
$m$\cite{lf2}. Therefore, we can treat the frictional coefficient
$\Gamma$ as a constant.

 From the total momentum conservation equation of the perturbed system
$\delta (h_{a}^{~b} T_{b~~;c}^{~c})=0$, one can identify
\begin{equation}
\psi_r=-\psi_{\phi} = \dot{\phi} \delta{\phi}.
\end{equation}

\subsection{Equations of perturbations}

Based on the perturbation variables given above, the evolution of 
these variables is governed by the following equations:
\begin{eqnarray}
&&\kappa=-3\dot{\varphi}+3H\alpha+\frac{k^2}{a^2}\chi~,  \\
&&H\kappa-\frac{k^2}{a^2}\varphi=
 -4\pi G\left(\varepsilon_r+\dot{\phi}\delta\dot{\phi}-
 \dot{\phi}^2\alpha+V,_{\phi}\delta\phi\right)~,  \\
&&\kappa-\frac{k^2}{a^2}\chi=
 -12\pi G\left(\psi_r-\dot{\phi}\delta\phi\right)~,  \\
&&\dot{\chi}+H\chi=\alpha+\varphi~,  \\
&&\dot{\kappa}+2H\kappa=\left(\frac{k^2}{a^2}-3\dot{H}\right)\alpha+
 4\pi G\left[(3\gamma-2)\varepsilon_r+4\dot{\phi}\delta\dot{\phi}
  -4\dot{\phi}^2\alpha-2V,_{\phi}\delta\phi \right]~,  \\
&&\dot{\varepsilon}_r+3\gamma H\varepsilon_r=
  \frac{k^2}{a^2}\psi_r+\gamma\rho_r(\kappa-3H\alpha)+
2\Gamma\dot{\phi}\delta\dot{\phi}-\Gamma\dot{\phi}^2\alpha~,  \\
&&\dot{\psi}_r+3H\psi_r=-\gamma\rho_r\alpha-(\gamma-1)\varepsilon_r-
\Gamma\dot{\phi}\delta\phi~, \\
&&\delta\ddot{\phi}+(3H+\Gamma)\delta\dot{\phi}+\left(\frac{k^2}{a^2}
  +V,_{\phi\phi}\right)\delta\phi=
  \dot{\phi}(\kappa+\dot{\alpha})-
  \left[(3H+\Gamma)\dot{\phi}+2V,_{\phi}\right]\alpha~,
\end{eqnarray}
where $k/a$ is the magnitude of the physical wave vector.

The equations (3.26) - (3.30) can be obtained from the ADM constraints, the
ADM propagation equation, and the Raychaudhuri equation as in \cite{h123},
but considering the total perturbations of energy, pressure and energy 
flux are given by
\begin{equation}
\varepsilon=\varepsilon_{\phi}+\varepsilon_{r}~,\hspace{6mm}
\pi=\pi_{\phi}+ \pi_{r}~,\hspace{6mm}
\psi=\psi_{\phi}+ \psi_{r}~,
\end{equation}
respectively.  Equations (3.31) and (3.32) are derived from the energy and
momentum equations of thermal bath, and Eq. (3.33) from the $\phi$ field
equation. 

Since the background 3-space is homogeneous and isotropic, all the
perturbation variables are gauge independent under purely spatial gauge
transformations. The perturbation equations presented above are also
independent of the spatial gauge transformation. However, they have no
physical meaning if the constant-time hypersurface they are embedded in is
not fixed \cite{b88}. To settle this issue, we use the uniform-curvature
gauge (UCG) \cite{h123} which is remarkably convenient in dealing with
relativistic perturbation analyses involving the scalar field. In this
case the set of perturbation equations can be obtained from Eqs. (3.26) to
(3.33) by setting up $\varphi\equiv 0$.  Furthermore, by virtue of the
definitions of $\varepsilon_{\phi},\ \pi_{\phi},\ \psi_{\phi}$, one can
establish the following relations
\begin{eqnarray}
\varepsilon_{\rm tot}&\equiv&\varepsilon_r+\varepsilon_{\phi}=
  3H\psi_{\rm tot}-\frac{H}{4\pi G}\frac{k^2}{a^2}\chi~,  \\
\psi_{\rm tot} &\equiv &\psi_r+\psi_{\phi}=
  -\frac{H}{4\pi G}(\dot{\chi}+H\chi)~.
\end{eqnarray}
These are connections between the matter variables and the metric variables. 

\section{The super-Hubble suppression}

\subsection{The initial conditions of perturbations}

In order to solve the evolution of the perturbations Eqs. (3.26) -(3.33),
the initial conditions to $\delta\phi$, $\delta\dot{\phi}$, $\chi$ and
$\dot{\chi}$ are needed.

It has been shown by using stochastic inflations\cite{rey,star} that
in the slow-rolling regime of warm inflation the long-wavelength
($k\ll H$) modes of the $\phi$ field perturbations satisfies\cite{lf2}
\begin{equation}
\frac{d\delta{\phi}}{dt} \simeq -
\frac{V,_{\phi \phi}+ k^2a^{-2}}{3H+\Gamma} \delta{\phi} + \eta.
\end{equation}
where the stochastic term $\eta$ caused by the high frequency
fluctuations satisfies
\begin{equation}
\langle \eta(t)  \rangle =0,  \hspace{1cm}
\langle \eta(t) \eta(t') \rangle
=\frac{H^2T}{2\pi}\delta(t-t').
\end{equation}

We are only interested in the period in which the physical
scale of the perturbations equal to, or large than the Hubble radius
$H^{-1}$, i.e. $aH/k \leq 1$. Moreover, the slow-roll condition
$|V,_{\phi\phi}| \ll 9H^2$, and therefore, Eq. (4.1) becomes
\begin{equation}
\frac{d \delta \phi}{dt}= - 
\frac{k^2a^{-2}}{3H+\Gamma}\delta \phi + \eta.
\end{equation}

At the initial time, i.e. the horizon-crossing moment $t$,
we have $ak/H \simeq 1$. Eq.(4.3) gives the initial 
fluctuations of $\phi$ field as
\begin{equation}
\langle \delta \phi \rangle
\simeq \left ( \frac{3}{4\pi}HT  \right )^{1/2}.
\end{equation}
The initial value of $\delta \dot{\phi}$ can be also estimated by
(4.3) as
\begin{equation}
\delta\dot{\phi} \simeq -
\frac{H^2}{3H+\Gamma} \delta{\phi}.
\end{equation}

The initial conditions for $\chi$ and $\dot{\chi}$ can be determined by
the relations (3.35) and (3.36), which give
\begin{eqnarray}
\chi~ & = & \frac{4\pi G}{H}\frac{a^2}{k^2}
\left[3H(\psi_r+\psi_{\phi})-(\varepsilon_r+\varepsilon_{\phi})\right], \\
\dot{\chi}~ & = & 
-\frac{4\pi G}{H}\left (\psi_r+\psi_{\phi} \right )-H\chi~.
\end{eqnarray}

We can estimate the initial fluctuations of the radiation component
by considering that it is in thermal equilibrium on sub-Hubble
scales. In this case, the thermal fluctuation is simply described by
$\delta\rho_r/\rho_r\simeq 1/\sqrt{n_r}$, where $n_r$ denotes the total
number of photons within the horizon $H^{-1}$. Because the photon number
density is proportional to $T^3$, and the volume inside the Hubble radius
is about $(4\pi/3)H^{-3}$, we have then
\begin{equation}
\frac{\delta\rho_r}{\rho_r} \simeq
 \sqrt{\frac{3}{4\pi}}\left(\frac{H}{T}\right)^{3/2}.
\end{equation}
As such, the energy fluctuations caused by $\delta\rho_r$ are given by
\begin{equation}
\delta\rho_r \sim
 \frac{\delta\rho_r}{\rho_r}T^4\simeq \sqrt{\frac{3}{4\pi}HT}\cdot
  HT^2=\delta\phi HT^2 = \delta\rho_{\phi} \frac {HT^2}{V'(\phi)}.
\end{equation}
In the case of weak dissipation, $\rho_r < \dot{\phi}^2/2$, or 
$HT^2 < V'(\phi)$, holds during the inflation epoch \cite{lf2}.  
Accordingly, we have
\begin{equation}
\varepsilon_r = \delta\rho_r \ll \delta\rho_{\phi}.
\end{equation}
Therefore, it is reasonable to approximate $\varepsilon_r\simeq 0$.

\subsection{Adiabatic and isocurvature perturbations}

The initial condition (4.10) implies that the radiation in the sub-Hubble
scales are almost isothermal. On the other hand, the $\phi$ field is
perturbed. Consequently, the equation of state is inhomogeneous, and 
therefore the perturbations produced by the above-mentioned initial
conditions generally are hybrid.

The energy perturbations of the $\phi$ field and the radiation component can
be decomposed into adiabatic (ad) and isocurvature (iso) modes as
\begin{eqnarray}
\varepsilon_{\phi} &=&
  \varepsilon_{\phi}({\rm ad}) + \varepsilon_{\phi}({\rm iso}),  \\
\varepsilon_r &=& \varepsilon_r({\rm ad}) + \varepsilon_r({\rm iso}).
\end{eqnarray}
By definition, the adiabatic perturbations $\delta_{ad}$ are given by
\begin{equation}
\delta_{ad}=\frac {\varepsilon_{\phi}({\rm ad})}{\rho_{\phi} +p_{\phi}}
=\frac{\varepsilon_r({\rm ad})}{\rho_r + p_r},
\end{equation}
while the isocurvature mode satisfies
\begin{equation}
\varepsilon_{\phi}({\rm iso}) + \varepsilon_r({\rm iso}) = 0.
\end{equation}
Therefore, the adiabatic perturbations can be rewritten as
\begin{equation}
\delta_{ad}\equiv\frac{\varepsilon_{\phi}+\varepsilon_r}
{\rho_{\phi} +p_{\phi} + \rho_r + p_r},
\end{equation}
and the entropic perturbations are characterized by
\begin{equation}
S_{\phi r}\equiv \frac{\varepsilon_{\phi}}{\rho_{\phi}+p_{\phi}}-
\frac{\varepsilon_r}{\rho_r+p_r}=
\frac{\varepsilon_{\phi}({\rm iso})}{\rho_{\phi}+p_{\phi}}-
\frac{\varepsilon_r({\rm iso})}{\rho_r+p_r}.
\end{equation}

Using the initial conditions Eq. (4.10) we have
\begin{equation}
\varepsilon_r({\rm ad}) = - \varepsilon_r({\rm iso}).
\end{equation}
Thus, Eqs. (4.13) - (4.17) yield
\begin{equation}
S_{\phi r}=\left (1+ \frac{\rho_r + p_r}{\rho_{\phi} +p_{\phi}} \right )
   \delta_{ad} = (R_{\phi r}+1) \delta_{ad}
\end{equation}
where
\begin{equation}
R_{\phi r} \equiv \frac{\rho_r + p_r}{\rho_{\phi} +p_{\phi}}.
\end{equation}
Accordingly, the initial ratio between  is
\begin{equation}
\frac{S_{\phi r}}{\delta_{ad}}= R_{\phi r} + 1 .
\end{equation}
Using the solutions (2.8) and (2.9) we have
\begin{equation}
R_{\phi r}= \frac{4}{3}\frac{\rho_r}{\dot{\phi}^2}
   \simeq \frac{2}{3}\frac{\Gamma}{H}.
\end{equation}
Therefore, the initial ratio between the amplitudes of adiabatic and
entropic perturbations is determined by the coupling coefficient $\Gamma$
between the $\phi$ field and the thermal bath. In the case of $\Gamma \gg
H$, the entropic perturbations are dominant. On the other hand, when
considering models with $\Gamma < H$, we always have $S_{\phi r} \simeq
\delta_{ad}$ regardless of the value of $\Gamma$.

\subsection{Evolution of the adiabatic and entropic perturbations}

Now we calculate the evolution of $\delta_{ad}$ and $S_{\phi r}$, which
are, respectively, the adiabatic and isocurvature modes of perturbations
when reentering the horizon. By virtue of Eqs. (3.26)-(3.33), it is
straightforward to obtain
\begin{equation}
\dot{\delta}_{ad}=\frac{k^2}{a^2}\left(\chi+\frac{\psi_{\phi}}
{\rho_{\phi}+p_{\phi}}\right) -
\left(\Gamma+2\frac{V,_{\phi}}{\dot{\phi}}\right)\alpha +
2\frac{V,_{\phi}}{\dot{\phi}^2}\left(\delta{\dot{\phi}} -
\frac{\ddot{\phi}}{\dot{\phi}}\delta\phi\right).
\end{equation}
\begin{eqnarray}
\dot{S}_{\phi r} & = & \frac{k^2}{a^2}\left(\frac{\psi_{\phi}}
{\rho_{\phi}+p_{\phi}}-\frac{\psi_r}{\rho_r+p_r}\right) -
\Gamma\left(\frac{\varepsilon_r}{\rho_r}\right) -  
2\left(\Gamma+\frac{V,_{\phi}}{\dot{\phi}}\right)\alpha \\ \nonumber
  & & + 2\frac{V,_{\phi}}{\dot{\phi}^2}\left(\delta{\dot{\phi}}-
\frac{\ddot{\phi}}{\dot{\phi}}\delta\phi\right) +
2\Gamma\left(\frac{\delta\dot{\phi}}{\dot{\phi}}\right).
\end{eqnarray}

In the super-Hubble regime, i.e. $k \ll aH$, terms containing $k^2$ and
$\delta\dot{\phi}$ in (4.22) and (4.23) are negligible.  Meanwhile, 
the slow-rolling of the inflation entails $\ddot{\phi} \simeq 0$.
Therefore, for the super-Hubble evolution, (4.22) and (4.23) reduce to,
respectively,
\begin{equation}
\dot{\delta}_{ad}\simeq
-\left(\Gamma+2\frac{V,_{\phi}}{\dot{\phi}}\right)\alpha,
\end{equation}
and 
\begin{equation}
\dot{S}_{\phi r}\simeq
-2\left(\Gamma+\frac{V,_{\phi}}{\dot{\phi}}\right)\alpha,
\end{equation}
where $\varepsilon_r \ll \rho_r$ has been used.
Owing to Eqs. (3.29) and (3.36), the perturbed lapse function $\alpha$
in the UCG is given by
\begin{equation}
\alpha=\dot{\chi}+H\chi=-\frac{4\pi G}{H}\psi_{\rm tot}.
\end{equation}
Thus, Eq. (3.25) assures $\alpha \simeq 0$ such that both $\delta_{ad}$
and $S_{\phi r}$ retain unchanged during the super-Hubble evolution.

A typical numerical solution of $\delta_{ad}$ and $S_{\phi r}$ is shown
in Fig. 2, which is given by a well-tested routine ODEX \cite{hnw}
with exercises of the standard numerical scheme (e.g. the appendices of
\cite{sbb,viana}). Figure 2 shows that the amplitude of $\delta_{ad}$
evolves as a constant when its scale crosses out of horizon. Even in
the range $t>t_f$, $\delta_{ad}$  still maintains constant until it
re-enters horizon. The constancy of
$S_{\phi r}$ in the period of $t_e <t<t_f$ can also been seen from Fig. 2. 

Evidently, the ratio $R_{\phi r}$ [Eq. (4.21)] given by the solutions
(2.8) and (2.9) is also a constant in the range $t>t_f$. Thus, Eqs. (4.20)
and (4.21) are valid not only at the initial time $t_x$, but also in the
whole regime of warm inflation. One can then conclude that at the time
$t_f$, i.e. the starting moment of the radiation dominated universe, the
ratio between the perturbations of entropic and adiabatic modes on
super-Hubble scales is entirely determined by the initial value of (4.20).

\subsection{Super-Hubble suppressed spectrum}

The super-Hubble suppressed power spectrum (1.2) can be approximately
rewritten as
\begin{equation}
P(k) \propto k^n f(k),
\end{equation}
and the function $f(k)$ is given by
\begin{equation}
f(k) = 1 - s \theta\left(\frac{k_0}{k}-1\right)
\end{equation}
where $\theta(x)$ is a step function and $k_0=2\pi H_0$ is the present
comoving wavenumber of the Hubble scale. The parameter $s$ represents the
super-Hubble suppression factor. For $k <k_0$, the density perturbations
are only given by adiabatic fluctuations, i.e. the amplitude of the power
spectrum $P(k)$ is proportional to $\delta^2_{ad}$. For $k>k_0$, however,
the density perturbations are given by both adiabatic and isocurvature
fluctuations.  If the isocurvature perturbations become density
perturbations of the same amplitude after re-entering the horizon, the
amplitude of the power spectrum $P(k)$ is proportional to $\delta^2_{ad} +
S^2_{\phi r}$. Thus, the super-Hubble suppression factor can be defined as
\begin{equation}
s\equiv \frac{S_{\phi r}^2}{S_{\phi r}^2 + \delta_{ad}^2}.
\end{equation}
Using the initial ratio Eqs.(4.20) and (4.21), we have
\begin{equation}
s \simeq \frac{[(2\Gamma/3H)+1]^2}{[(2\Gamma/3H)+1]^2 +1}
\end{equation}
Therefore, the suppression factor $s$ is at least 0.5, regardless of
$\Gamma$. Figure 3 shows $s$ as a function of $\Gamma/H$.

\section{Conclusions and Discussion}

We investigated the evolution of the primordial density perturbations
produced by the inflation with thermal dissipation. A full relativistic
calculation of the evolution of the perturbations from the era of warm
inflation to a radiation dominated universe has been developed. The
emphasis is on tracking the ratio between the adiabatic and the isocurvature
mode of the initial perturbations. This result is employed to explain
the suppression of the cosmic temperature fluctuations on scales larger than
Hubble radius.

We showed that the super-Hubble suppression factor $s$ for a warm inflation
scenario is mainly dependent on the friction parameter $\Gamma$. More
interestingly, we found that $s$ is at least $0.5$ regardless of the
details of the parameters. This prediction would be a serious test for the
model of inflation with thermal dissipation.

\acknowledgments

We thank Andrew Engel and Mary Domm for their comments on the manuscript.
W.L.L. would like to thank Pedro Viana and John Cocke for helpful
discussions at the early stage of this work.  He also acknowledge the
support from the National Science Council, R.O.C. via Grants
No. NSC89-2112-M-001-001.


\begin{figure}
\caption{A typical solution of the background evolution for the inflation 
with thermal dissipation. The parameters of the model are taken to be 
$M=10^{15}$ GeV, $\lambda=1.2\times 10^{-16}$, and $\Gamma=H\ $. All
parameters shown here have been de-dimensionalized by $H$. The Hubble
parameter $H$ remains constant during the slow roll period. The energy
density $\rho_r$ and the temperature $T$ of radiation (thermal bath) increase
with time, and smoothly enter into a radiation dominated regime at $t_f$. The
kinetic energy of $\phi$ field $\mbox{KE}_{\phi}=\dot{\phi}^2/2$ increases
with time also.
The warm inflation era is from $t_e$ to $t_f$, thus the thermal
duration $N\equiv H(t_f-t_e)\simeq 72$ in this example. 
}
\label{1}
\end{figure}

\begin{figure}
\caption{A typical solution of the evolution of the adiabatic mode 
$\delta_{ad}$ and isocurvature mode $S_{\phi r}$ of the primordial 
perturbations during warm inflation. The dot-dashed line represents the 
amplitude of the adiabatic mode $\delta_{ad}^2$. The solid line denotes the 
amplitude of the isocurvature mode $S_{\phi r}^2$.  Both of them become
constant when they cross out of the Hubble horizon at $t_x$.  The
amplitude of the adiabatic spectrum $\delta_{ad}^2$ remains constant even
after the inflation when $t > t_f$.} 
\label{2} 
\end{figure}

\begin{figure}
\caption{The super-Hubble suppression factor $s$ as a function of
parameter $\Gamma=\Gamma(H)$. Evidently, $s > 0.5$ as $\Gamma \geq H$. 
}
\label{3}
\end{figure}

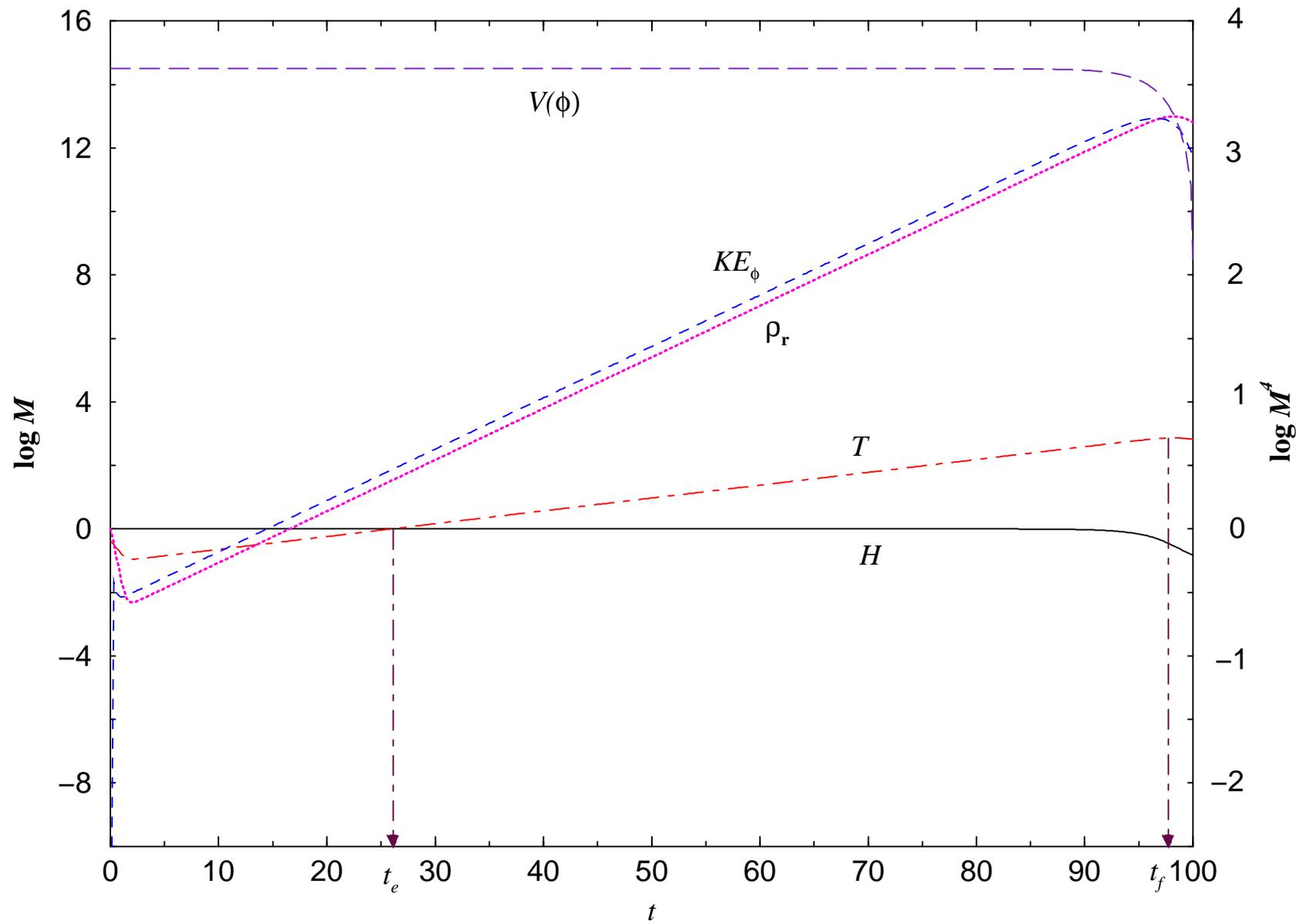
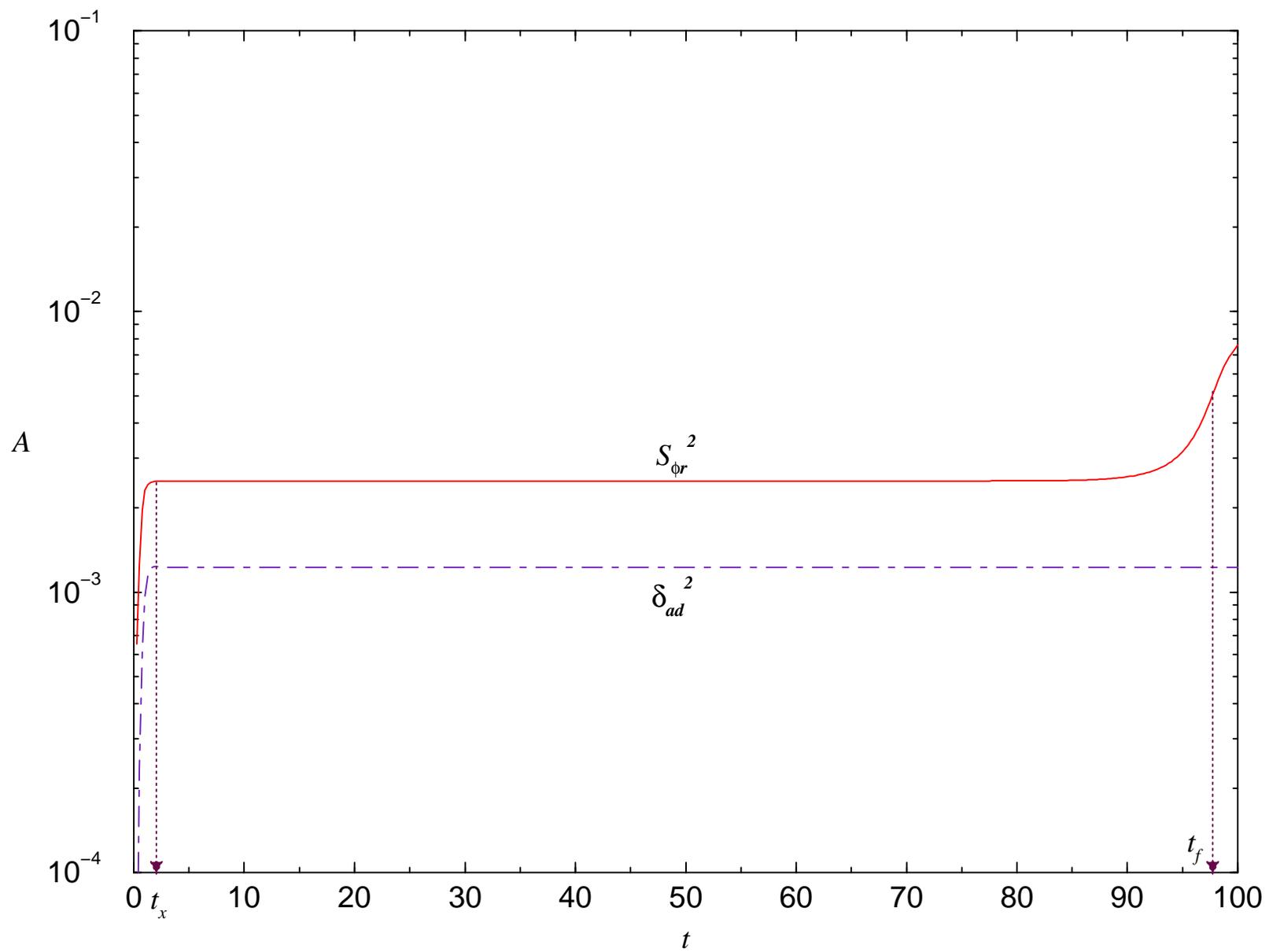
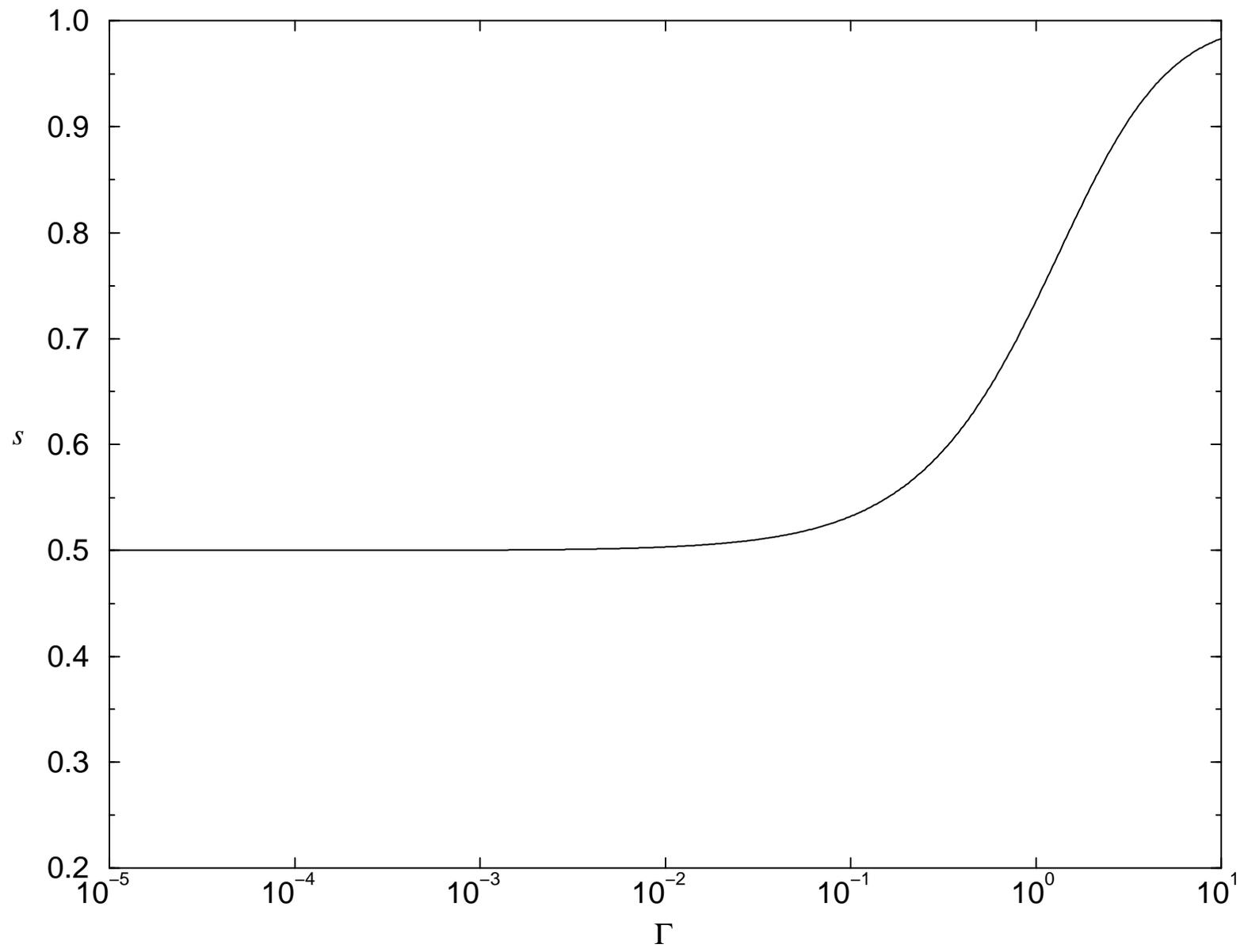
\end{document}